\documentclass[preprint2]{aastex6}
\usepackage{longtable,hyperref}
\pdfoutput=1
\newcommand{\beq}{\begin{equation}}
\newcommand{\eeq}{\end{equation}}

\begin{document}

\title{Modeling the Performance of the LSST in Surveying the Near-Earth Object Population}

\shorttitle{LSST and the NEO Population}
\shortauthors{Grav {\it et al.}}
\medskip

\author{T.~Grav} 
\affil{Planetary Science Institute, Tucson, AZ 85719, USA; tgrav@psi.edu}
\author{A.~K.~Mainzer} 
\affil{Jet Propulsion Laboratory, California Institute of Technology, Pasadena, CA 91109, USA}
\author{T.~Spahr} 
\affil{NEO Sciences, LLC}

\date{\rule{0mm}{0mm}}
\begin{abstract}
We have performed a detailed survey simulation of the LSST performance with regards to near-Earth objects (NEOs) using the project's current baseline cadence. The survey shows that if the project is able to reliably generate linked sets of positions and times (a so-called ``tracklet") using two detections of a given object per night and can link these tracklets into a track with a minimum of 3 tracklets covering more than a $\sim12$ day length-of-arc, they would be able to discover $62\%$ of the potentially hazardous asteroids (PHAs) larger than $140$ m in its projected 10 year survey lifetime. This completeness would be reduced to $58\%$ if the project is unable to implement a pipeline using the two detection cadence and has to adopt the four detection cadence more commonly used by existing NEO surveys. When including the estimated performance from the current operating surveys, assuming these would continue running until the start of LSST and perhaps beyond, the completeness fraction for PHAs larger than $140$ m would be $73\%$ for the baseline cadence and $71\%$ for the four detection cadence. This result is a lower completeness than the estimate of \citet{Ivezic.2007a} and \citet{Ivezic.2008a}; however the result is quite close to that of \citet{Jones.2016a} who show completeness $\sim70$\% using the identical survey cadence as used here. We show that the traditional method of using absolute magnitude $H < 22$ mag as a proxy for the population with diameters larger than $140$m results in completeness values that are too high by $\sim5$\%. Our simulation makes use of the most recent models of the physical and orbital properties of the NEO and PHA populations, as well as simulated cadences and telescope performance estimates provided by the LSST project.  The consistency of the results presented here when compared to those of \citet{Jones.2016a} demonstrates the robustness of these survey modeling approaches.  We also show that while neither LSST nor a space-based IR platform like NEOCam individually can complete the survey for $140m$ diameter NEOs, the combination of these systems can achieve that goal after a decade of observation.  

\end{abstract}
\section{Introduction}

As Earth travels through space along its orbit around the Sun it traverses through a population of asteroids and comets, termed the near-Earth objects (NEOs; asteroids and comets that approach within 1.3 au of the Sun). With regularity, Earth finds itself on a collisional course with one of these NEOs. Most of the time the objects are small and burn up in the atmosphere, but sometimes the NEO is large enough that it can cause damage. The effects can range from the local damage caused by the Tunguska airburst in 1908 \citep{Chyba.1993a} or the Chelyabinsk airburst in 2013 \citep{Brown.2013a,Kohout.2014a}, to the regional destruction caused by the impact that created the kilometer-sized Meteor Crater in Arizona \citep{Grieve.1987a}. There is even evidence that the effects can be global, as seen in the theory that an impactor in the Cretaceous-Paleogene era caused the enormous ring-shaped feature beneath the Gulf of Mexico and significantly contributed to the extinction of the dinosaurs \citep{Alvarez.1980a}. 

Over the last few decades, Congess has tasked NASA with two goals to address the NEO detection problem, in part spurred by the impact of Comet Shoemaker-Levy 9 into Jupiter \citep{Hammel.1995a,Zahnle.1994a}. The first goal, also called the Spaceguard survey, concerned the population of objects larger than 1km in diameter and directed the agency to detect $90\%$ of this population by 2008. \citet{Mainzer.2011e} showed that this goal was reached somewhere in 2010. The second goal was given in the George E. Brown, Jr. near-Earth object section of the NASA Authorization Act of 2005 (Public Law 109-155), which charged the agency to discover and track $90\%$ of the NEO population with diameters larger than $140$m by 2020.

The current NEO survey capability is dominated by the Catalina Sky Survey \citep[CSS;][]{Larson.2007a} and the Pan-STARRS project \citep{Denneau.2013a}. Both surveys operate in similar fashion by observing each pointing on the sky 4-5 times per night, with each return to the same pointing being separated by minutes to tens of minutes. Detections in the fields are connected into tracklets, and when their motions are consistent with that of the NEO population they are posted as candidate NEOs on the Minor Planet Center (MPC) NEO confirmation page. In many cases other observers, dubbed the follow-up community, provide further observations over the next few nights that refine the orbit and establishe whether or not the candidate was indeed an NEO. Originally the Pan-STARRS project planned to adopt a two detections per night (hereafter 2-detection) survey cadence for its wide-field multi-science survey. Due to a variety of problems, the Pan-STARRS project was forced to change its survey cadence to the more traditional four detections per night (hereafter 4-detection) cadence that has proven to be successful in CSS, NEOWISE \citep{Mainzer.2011a}, and now Pan-STARRS. See \citet{Denneau.2013a} for a description of the Pan-STARRS Moving Object Processing System (MOPS), including the details of its performance and its problems. New cameras, such as the Dark Energy Camera \citep[DECam;][]{Diehl.2012a}, together with refined false detection handling through for example machine learning algorithms \citet[see ][]{Goldstein.2015a} may yet provide the advances to solve this problem in the future (although at present these techniques are not yet widely implemented for NEO surveys).

The known NEO population currently consists of more than 13,800 objects, $\sim1,500$ of which were discovered in 2015 (http://neo.jpl.nasa.gov/stats/). The Catalina Sky Survey has been leading the NEO survey effort during most of the last decade, but when Pan-STARRS went from a multi-science survey telescope to a dedicated NEO survey telescope in early 2014, its discovery rate increased by $60\%$. This elevated Pan-STARRS to the premier discovery telescope in 2015, responsible for $48.6\%$ of the discoveries for that year. CSS was the second largest discoverer in 2015, responsible for $36.2\%$ of the discoveries. However, some caution must be exercised when considering where the increase in discoveries took place. The increase in the yearly discovery rate from 2011 to 2015 has been $74.2\%$, with 1,563 discoveries in 2015 compared to 897 in 2011. However, the increase in the yearly discovery rate of the larger objects (absolute magnitude $H < 22$ mag, traditionally considered to be objects larger than $140$m in diameter) is only $34.5\%$ over the same time period, with 526 discovered in 2015 compared to 391 in 2011. The fraction of objects discovered per year that have $H > 22$ mag has steadily increased from $56.4\%$ in 2011 to $67.6\%$ in 2015. This trend continues into the first two months on 2016. 

In 2003 NASA commissioned a report that concluded that a number of ground-based, space-based, and networked systems are capable of meeting the goal set forth in the George E. Brown, Jr. goal \citep{Stokes.2003a}. The National Research Council released a report in 2010 that came to a similar conclusion \citep{nrc_pd_report2010a}. Several projects have been proposed to address this problem, but detailed simulations of most of these proposals have not been published in the refereed literature. We performed a detailed study of an infrared space-based option in \citet{Mainzer.2015a}, where we compared the completeness for NEO population with diameters larger than $140$m for a Venus-trailing orbit and a L1 halo orbit. Sentinel, a privately funded infrared space-based survey, proposed by the B612 Foundation \citep{Lu.2013a}, is similar to the Venus-trailing orbit studied in \citet{Mainzer.2015a}. The L1 halo orbit option, named the Near-Earth Object Camera (NEOCam), has been been submitted to the 2005, 2010, and 2014 NASA Discovery Announcements. NEOCam has received funding in 2010 to continue the development of its infrared detector technology \citep{McMurtry.2013a} and was selected for Phase 2 development in the 2014 Discovery Announcement Opportunity. \citet{Mainzer.2015a} concluded that the L1 halo option offered superior performance to the Venus-trailing option when considering integral survey completeness for NEOs larger than $140$m in diameter, even when the effects of the loss of data rate in a Venus-trailing orbit is neglected.

In this paper we examine a complementary ground based option, the Large Synoptic Survey Telescope \citep[LSST;][]{Ivezic.2007a,Ivezic.2008a}, an 8.4m telescope (with the approximate effective collective area of a 6.67m in diameter mirror) with a 9.6 square degree imager that is currently being built in Chile with funding from the National Science Foundation and the Department of Energy. We perform detailed survey simulations using the latest population models based on NEOWISE results \citep{Mainzer.2011e,Mainzer.2012b} and the most current model of the survey plan provided by LSST project and compare our results with others published by the LSST project  \citep{Ivezic.2007a,Ivezic.2008a}. These results are compared to the best available results from the space-based L1 halo orbit survey option. We also show that assuming that using absolute magnitude $H < 22$ mag as a proxy for the population with diameters larger than $140$m is not a valid approximation. 

\subsection{The LSST Project}

The effectiveness all surveys for NEO discovery depends on the details of their performance and observational cadence. As the baseline LSST survey cadence is currently envisioned, each pointing will be visited two time per night. This will produce a 6 band ($u$, $g$, $r$, $i$, $z$, and $y$ filters) wide-field deep astronomical survey covering more than 20,000 square degrees of southern sky, visiting each pointing on the sky over 1000 times in a 10 year survey.

In \citet{Ivezic.2007a}, the LSST team presented a simulation that used a set of 1000 synthetic orbital elements that match the distribution of discovered objects in the large size range where present surveys were essentially complete. The simulation computed the positions of the synthetic orbits every 5 days and used a filtering method based on the assumed sky coverage and cadence pattern, limiting magnitude of the survey, visibility constraints, and observing conditions. They estimated that the LSST baseline cadence can achieve, over 10 years, a completeness of $90\%$ for PHAs larger than $250$m in diameter, and $75\%$ completeness for those with diameters larger than $140$m. They further suggested that ongoing simulations showed that improvements in filter choices and operations, LSST would be capable of reaching a completeness of $90\%$ for PHAs larger than $140$m in ten years, but details of these optimizations were not discussed. In \citet{Ivezic.2008a} a size-limited complete sample of 800 known PHAs was used as the trial population. This simulation improved upon the previous simulation by determining which PHAs are present in each exposure and whether they were bright enough to be detected in individual exposures. They found that the baseline cadence provides orbits for $82\%$ of the PHAs larger than $140$m in diameter after 10 years of operations. It was suggested that optimization of certain aspects of the survey operations would improve the completeness to $90\%$ in 12 years without significantly affecting the survey's other science goals. It is important here to note that sizes quoted in these papers are derived by converting absolute magnitudes using an average albedo. We will show in our results why this is not an optimal approach, resulting in completeness values that are overly optimistic. 

\section{The Simulations}

In the last few years our knowledge of the NEO population has increased significantly. \citet{Mainzer.2011e,Mainzer.2012b} used the asteroid-detection portion of the Wide-field Infrared Survey Explorer's \citep[WISE;][]{Wright.2010a} data processing pipeline \citep[NEOWISE;][]{Mainzer.2011a} to update the estimate of the number of objects larger than $140$m in diameter and to derive new size and albedo distributions for the NEO population as a whole, as well as the subpopulations. This has allowed us to create a new synthetic NEO population that is based on both the known NEO orbital population \citep{Grav.2011a} and these new distributions of sizes and albedos (see Section \ref{sec:synth_pop} for the details). Significant work has also been done by the LSST project, resulting in the production of an improved baseline cadence that features the latest knowledge of the telescope and instrument performance, weather and seeing conditions at the Cerro Pach\'on site, and improvements based on earlier cadence simulations. With these recent developments, we present a set of survey simulations that combine our updated NEO population model with the most recent cadence and performance simulations provided by the LSST project to investigate the performance of the LSST project in surveying the NEO population. 

\subsection{2 versus 4 Detections per Night}

One of the major features of the LSST baseline cadence is the use of the 2-detection approach, which deviates from the way the current NEO surveys operate. CSS, PanSTARRS and NEOWISE all require 4-5 detections per night to reliably link detections of individuals object due to the large range of rates and directions of most NEOs.  This number of repeated observations per night has been shown to be relatively robust against the noise points, image artifacts, cosmic rays, and other transient sources that degrade the reliability of position-time pairs (so-called {\it ``tracklets"}) constructed from fewer detections per night \citep{Denneau.2013a, Mainzer.2011a, Cutri.2012a}.

Since the baseline assumes a 2-detection cadence that has not yet been tested and validated for NEO discovery, the LSST project has also produced a 4-detection cadence, which is more similar to the cadences used by the current NEO surveys. These two cadences are described in more detail in Section \ref{sec:lsst_pattern}. 

In \citet{Mainzer.2015a}, we assumed that 4 detections spaced over 8-9 hours were required to reliably link detections of an individual object to form a tracklet. Requiring additional detections dramatically increases tracklet reliability but decreases the rate at which fresh sky can be covered, so the performance of 2-detection surveys cannot be compared with those of 4-detection surveys.  Should a 2-detection cadence be proven a viable means of linking tracklets through future testing and validation, space-based surveys can also adopt this approach. Therefore, to compare the relative performance of the LSST survey and the L1 halo-orbit infrared space-based survey in discovering and tracking NEOs, we have elected to simulate both the 2- and 4-detection cadences. This work provides the bounding cases for the expected performance of these two options and allows accurate comparisons to be made.

\subsection{The Solar System Survey Simulator}

The solar system survey simulator we have created to analyze the performance of LSST is essentially the same as that used to model the performance of a 0.5m thermal infrared space telescope in \citet{Mainzer.2015a}.  The technique begins by combining a frame-by-frame pointing list for the simulated survey with a population of synthetic moving objects whose positions and velocities are computed at the epoch of each frame.  The brightness and on-sky velocity of each object as it appears in each frame is evaluated to determine whether or not it would have been detected in that exposure, depending on that frame's estimated sensitivity.  If sets of detections are found in a cadence that allows them to be uniquely linked to one another over a sufficiently long timespan, the object can be declared {\it ``discovered"}.  The survey cadence is critically important in determining which objects merely pass through the field of regard, and which objects are actually detected, discovered, and tracked.  This simulation does not yet include models of background sources or other image artifacts and transient sources (such as noise, cosmic rays, defective pixels, scattered light, etc.) that can confuse or break linkages; therefore, these simulations should be regarded as a best case.

If a set of two 4-detection tracklets covering an average of 10-20 nights or more can be linked to one another, an asteroid's orbit can be determined with sufficient accuracy to allow it to not only be declared discovered, but to allow it to be recovered at its next apparition. A set of at least three 2-detection tracklets is needed to allow the same for the 2-detection cadence. The survey simulation in \citet{Mainzer.2015a} describes a method for tallying detections over a survey cadence that collects tracklets spanning $\sim$10-22 days.   In this paper, we apply a similar technique to simulate the performance of the LSST project, adopting the baseline survey cadence given on the LSST Operations Simulation (OpSim) website\footnote[1]{https://confluence.lsstcorp.org/display/SIM/Operations+Simulator +Benchmark+Surveys} \citep{Connolly.2014a, Jones.2014a}. This frame-by-frame simulation is combined with our synthetic population model to predict the numbers of objects that would be detected in each frame.  

To give a robust estimate of the variation in orbital elements and physical properties, our simulations include 25 synthetic populations generated randomly according to the size distribution, albedo distribution, and numbers specified in \citet{Mainzer.2012b}.  By running many populations through the survey simulator, we can evaluate systematic uncertainties introduced by the limits of our knowledge of the NEO population.  The ephemeris for each synthetic object was computed at each frame time using the SWIFT numerical integrator \citep{Levison.1994a}, which implements the Bulirsch-Stoer integration method, on the high performance computing facilities at Jet Propulsion Laboratory.  Objects were assumed to be successfully assembled into tracklets if they were detected two or more times per night, with an on-sky velocity between 0.011 and 3.2$^{\circ}$/day.  The slow speed limit of $\sim$0.011$^{\circ}$/hour is set by the LSST average seeing of $\sim$0.5 - 1 arcsec and the minimum estimated separation between two consecutive exposures in a night of $\sim$30 minutes.  The upper speed limit of is determined by the need to avoid significant trailing losses in a 15 sec exposure. 

We note here that these velocity limits have virtually no effect on the completeness fraction of these survey simulations as we focus on the population larger than $140$m. For these larger objects the distances at which they are observed are such that their velocities naturally fall within these limits. However, for objects smaller than $140$m, the effects of trailing can have a significant impact on the completeness fraction as these objects have to be much closer to the observatory to be detectable and thus generally have much higher velocities with respect to the observer. For the baseline cadence, tracklets were considered successfully linked if three tracklets, each containing at least two detections each, were detected over the course of $\sim$12 days. This requires that a single 2-detection tracklet can be successfully linked to another 2-detection tracklet no later than 6 days after, which as noted above has not been yet proven to be workable to date; we do not address this issue at present in our simulation.  For the 4-detection cadence, an object was considered a discovery if two tracklets, each with at least four detections, are found within a 12 day timespan. This timespan was selected from our experience with the current surveys, where follow-up and linking has proven to be extremely difficult after 10-12 days, based on a 4-5 detection discovery tracklet \citep{Cutri.2012a, Mainzer.2011a, Denneau.2013a}.

As noted in \citet{Mainzer.2015a}, any survey simulation must account for the performance of the existing surveys such as the CSS and PanSTARRS, both in terms of the number of objects they have already discovered to date, and in terms of the number of objects they would be expected to discover over the course of the future survey.  It should be noted that the LSST survey patterns available from the OpSim start in 1994 and run 10 years from this date.  To estimate the overlap in discoveries between LSST and the historical and current surveys, it was therefore necessary to shift our historical and current survey simulations by 28 years back in time or shift the LSST cadence 28 years into the future (assuming that the LSST survey will start in 2022). This allowed us to derive a known population of NEOs that is similar to that which LSST will have when its starts its survey late this decade. Both methods yielded similar completeness fractions and similar overlap fractions with the existing surveys.

\subsection{The Synthetic Near-Earth Object Population}
\label{sec:synth_pop}

The NEO population has historically been divided into several subpopulations. Starting from the outside and going inward, we consider four separate sub-populations in this paper: 1) the Amors, objects with perihelion distance in the range $1.017 < q \leq 1.3$ au; 2) the Apollos, which have orbits with semi-major axis $a \geq 1$ au and perihelion distance $q \leq 1.017$ au; 3) the Atens, which have orbits with semi-major axis $a < 1$ au and aphelion distance $Q \geq 0.983$ au; and 4) the interior Earth objects (IEOs), which have aphelion distances $Q < 0.983$ au.  In addition, we use the term potentially hazardous asteroid (PHA) to indicate objects that are larger than $140$m (the definition is usually given as absolute magnitude $H < 22$ mag rather than size) and have minimum-orbital-intersection-distance (MOID) less than $0.05$ au \citep{Ostro.2004a}. 

\citet{Mainzer.2011e} determined based on the results of the NEOWISE survey that there are $20,500\pm3,000$ near-Earth asteroids (NEAs; near-Earth comets were not included in the study) larger than 100 m in diameter. They looked at the subpopulations of the NEA population in \citet{Mainzer.2012b} using the same data and determined that there are $1,600\pm760$ Atens larger than 100 m in diameter. The corresponding numbers for the Apollo and Amor subpopulations are $11,200\pm2,900$ and $7,700\pm3200$, respectively. This is similar to the results of \citet{Greenstreet.2012a}, which is based on numerical simulations of dynamical evolution of objects from the main asteroid belt source regions into NEO space. Since NEOWISE does not provide an estimate of the IEOs, we use the result of \citet{Greenstreet.2012a} as the basis to generate a population of interior objects that consist of $350\pm100$ objects (which is close to the $1.6\%$ fraction given in their paper).  The orbital elements for the Atens, Apollos, and Amors were generated based on the \citet{Grav.2011a} synthetic solar system model, which is in turn based on the \citet{Bottke.2002a} model.  

In \citet{Mainzer.2012b} each of the subpopulations was found to have a slightly different size and albedo distribution, and our synthetic populations reflect these differences. Since no measure of the size or albedo distribution of the IEOs exists, we use the distributions for the NEOWISE Aten population to generate sizes and albedos for the IEO population. 

Since the LSST survey collects exposures in each of six bands, we need to model the magnitude of the synthetic NEOs at each wavelength.  The absolute magnitude $H$ was found for each object using the relation \begin{equation}H = -5\cdot log_{10}\left( \frac{D \sqrt{p_{v}}}{1329}\right),  \end{equation}where $D$ is the assumed diameter in km and $p_{V}$ is the assumed geometric visible albedo \citep{Bowell.1989a}.  Each object's $V$ band magnitude was computed using the IAU phase curve correction: \begin{equation}V(\alpha) = H + 5 log(R \Delta) - 2.5 log[(1 - G)\Phi_{1}(\alpha) + G\Phi_{2}(\alpha)], \end{equation} where $R$ is the heliocentric distance in AU, $\Delta$ is the geocentric distance in AU, $\alpha$ is the Sun-observer-object angle, $G$ is the magnitude-phase relationship slope parameter, and $\Phi_{i}$ are given in \citet{Bowell.1989a}.  The magnitude in the $g$ band, $H_g$, is converted from the $V$ band by $g = V + 0.56(B-V) - 0.12$, where $B-V = (g-r+0.23)/1.05$ \citep{Fukugita.1996a}. Colors were generated in the $ugriz$ system using the color distributions found in \citet{Ivezic.2001a}, assigning C-type and S-type to each synthetic asteroid among the NEOs. While there are of course several other taxonomic classes found in the NEO population, the vast majority of objects are consistent with the two major classes \citep{Binzel.2002a, Mainzer.2011e}.

\subsection{LSST Survey Patterns}
\label{sec:lsst_pattern}
The two survey patterns used in this paper were generated by the LSST team and are available at their website. The most recent version of the baseline cadence is labeled {\it enigma\_1189}. The baseline cadence executes 5 science proposals: 1) the Wide-Fast-Deep survey, which is the universal cadence that covers large fractions (about 75\%) of the sky in pairs of exposures taken 15 to 60 minutes apart; 2) the Galactic Plane survey, which collects 30 visits in each of the 6 bandpasses; 3) the North Ecliptic survey, which covers the ecliptic in the universal cadence beyond the +20$^{\circ}$ declination limit set in the Wide-Fast-Deep survey; 4) the South Pole survey, which like the Galactic plane survey collects 30 visits in each bandpass; and 6) 6 deep drilling fields for supernova surveying. The baseline cadence starts and stops observing at 12$^{\circ}$ twilight and uses the CTIO 4m weather log as its weather model. It uses the latest telescope model and includes scheduled downtime for maintenance; slews and filter changes together take 6.4 sec on average.  The baseline cadence only uses the $u$ filter approximately 6 days per lunation. The 4-detection cadence is labeled {\it enigma\_1266}. This survey cadence follows the same design and criteria as the baseline, with the only significant difference being that each pointing is observed four times if possible. 

\begin{figure}[h]
\begin{center}
\includegraphics[width=8cm]{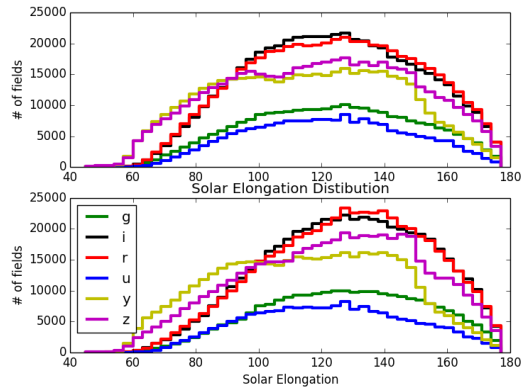}
\caption{The solar elongation distribution of the LSST cadences in each filter. The baseline 2-detection cadence is given in the top panel, while the 4-detection cadence is shown in the lower panel. }
\label{fig:LSST_solarelong}
\end{center}
\end{figure}

\begin{figure}[h]
\begin{center}
\includegraphics[width=8cm]{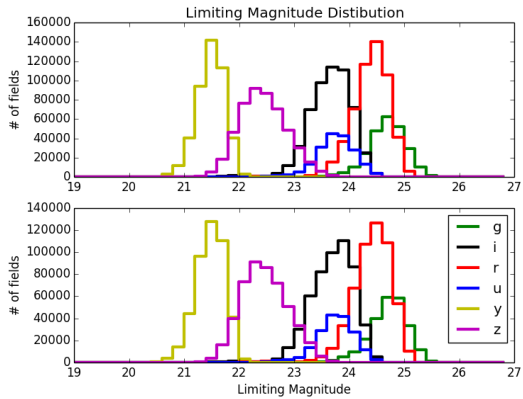}
\caption{The $5\sigma$ limiting magnitude distribution of the LSST cadences in each filter. The baseline 2-detection cadence is given in the top panel, while the 4-detection cadence is shown in the lower panel.}
\label{fig:LSST_limmag}
\end{center}
\end{figure}

\begin{figure}[h]
\begin{center}
\includegraphics[width=8cm]{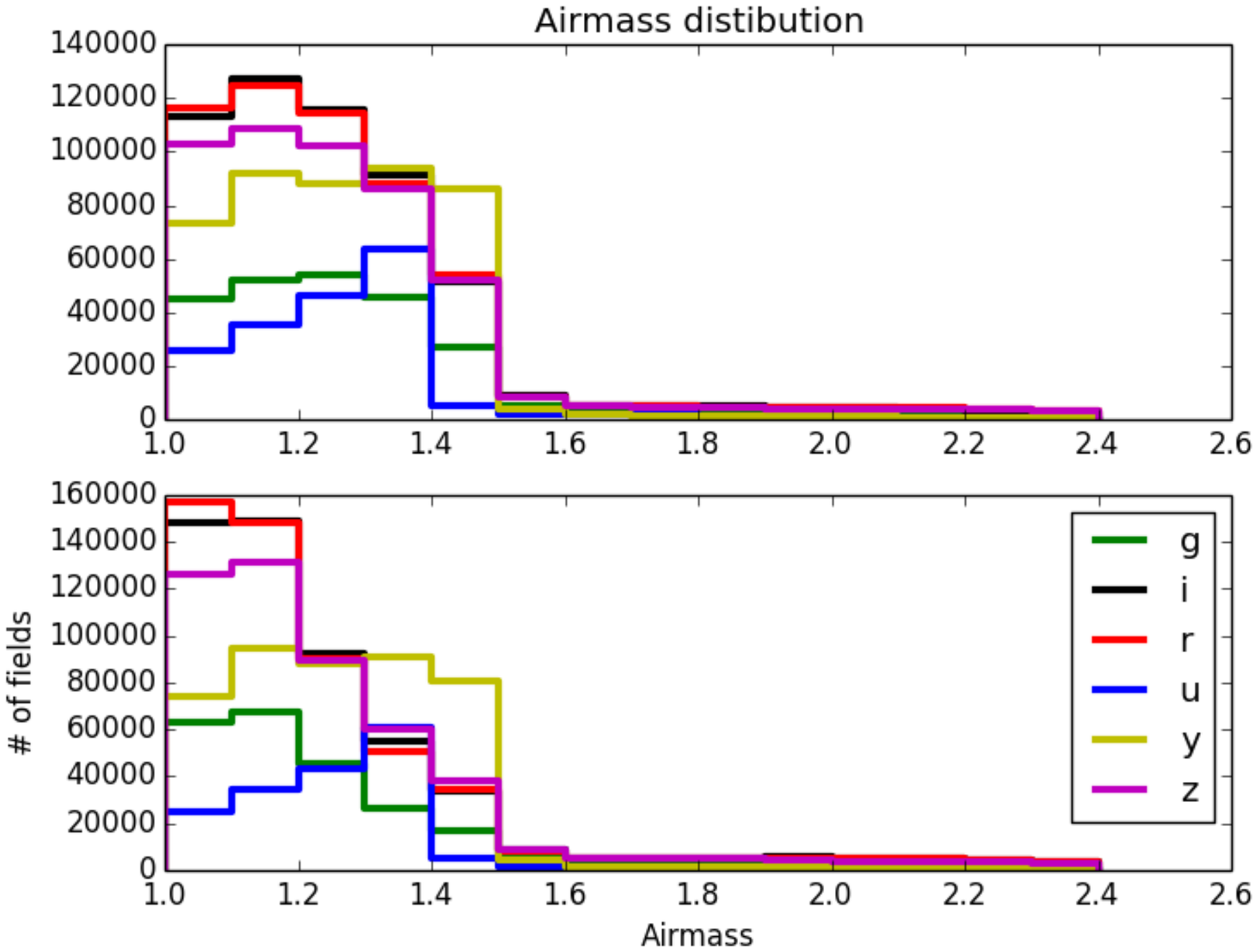}
\caption{The airmass distribution of the LSST cadences in each filter. The baseline 2-detection cadence is given in the top panel, while the 4-detection cadence is shown in the lower panel.}
\label{fig:LSST_airmass}
\end{center}
\end{figure}

\begin{figure}[h]
\begin{center}
\includegraphics[width=8cm]{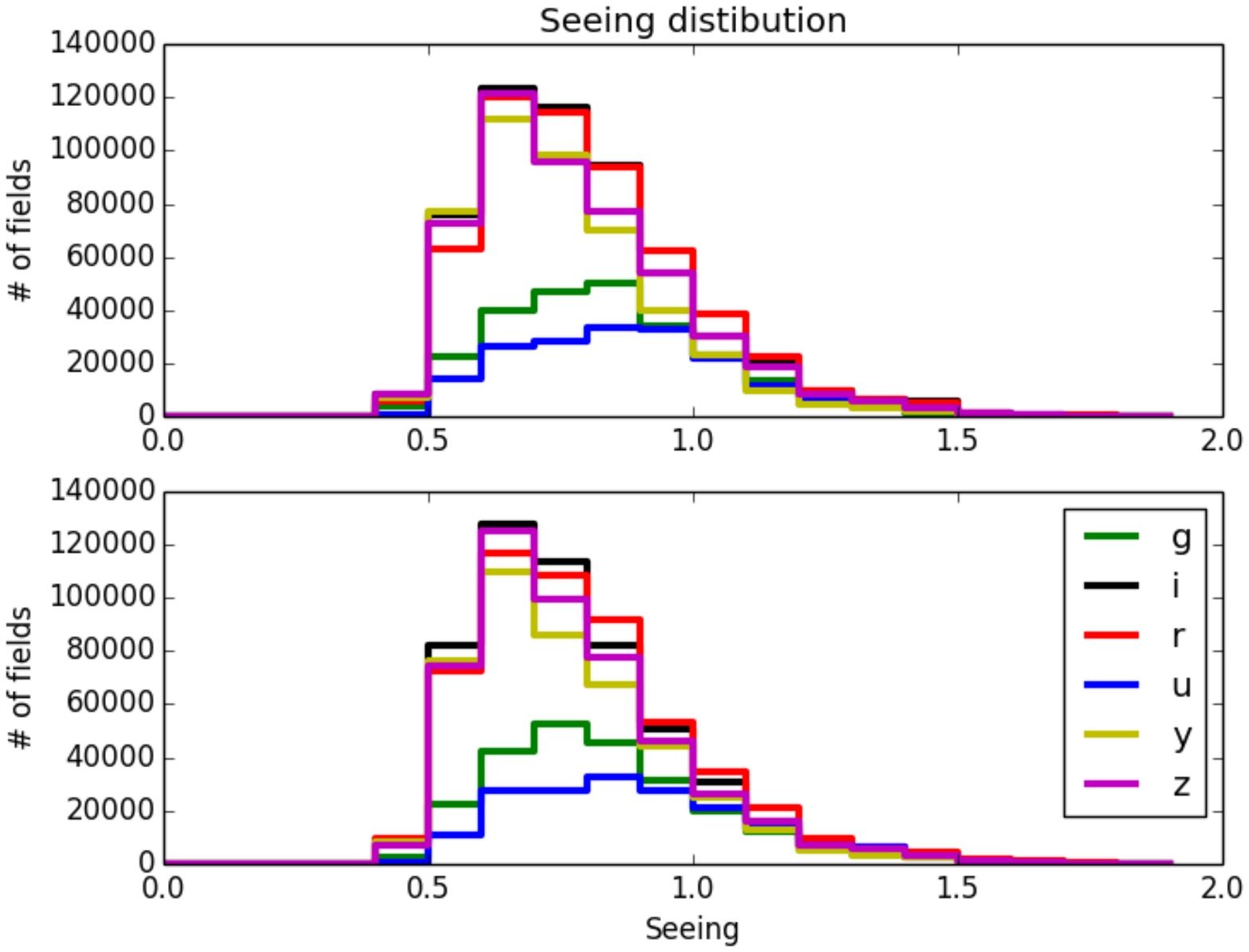}
\caption{The 10 year seeing distribution of the LSST cadences in each filter. The baseline 2-detection cadence is given in the top panel, while the 4-detection cadence is shown in the lower panel. }
\label{fig:LSST_seeing}
\end{center}
\end{figure}

The two LSST cadences are represented in Figures \ref{fig:LSST_solarelong} to \ref{fig:LSST_seeing}. The surveys cover solar elongations from opposition (at $180^\circ$) to about $60^\circ$, but only about 11\% of the observations are at solar elongations lower than $90^\circ$ for the baseline cadence (see Figure \ref{fig:LSST_solarelong}). This number drops to $\sim 7\%$ for the 4-detection cadence. Such limited coverage of these low-elongation regions will severely limit the usefulness of LSST in detecting objects in the IEO population, as these objects never rise above $90^\circ$ solar elongation.  The $5\sigma$ limiting magnitudes given by the published simulated survey cadences in each filter are very similar for the two cadences, with mean limiting magnitudes in each filter being $23.7$, $24.7$, $24.4$, $23.7$, $22.5$ and $21.5$ (for {\it u}, {\it g}, {\it r}, {\it i}, {\it z} and {\it y}, respectively).  It should be noted that though the seeing distributions are similar for the two cadences, the airmass distributions are not. The 2-detection baseline cadence spends significantly more time at higher airmasses, which is most likely due to increased time spent at lower solar elongations and the fact that more sky is covered each night. Therefore, more of the observations are made at less favorable airmasses. We note here that these $5\sigma$ limiting magnitudes are similar to those given in \citet{Ivezic.2008a}, but with lower values in the $z$ and $y$ filters, which we assume are due to the project having more recent values for the performance and observing conditions when these filters are used. The limiting magnitudes and solar elongation coverage presented here are at odds with \citet{Myhrvold.2016a}; his Figure 7 shows much fainter limiting magnitudes than presented here.  While \citet{Myhrvold.2016a} cites the possibility and capability of LSST observing at much smaller solar elongations, this is inconsistent with the published cadences provided by the LSST project and would potentially interfere with its other science goals. Additionally, \citet{Myhrvold.2016a} incorrectly assumes that an object need only be detected once for it to be counted as discovered, cataloged, and tracked. 

 \subsection{Detector Gaps}
LSST aims to have less than $5\%$ of the focal plane lost due to gaps, bad pixels, or dead detectors. With a pixel size of 10 microns, which equals 0.2 arc seconds of sky for the LSST telescope, we assumed gaps that were 1 mm wide, or equivalent to 100 pixels. There are 21 ``rafts" in the LSST focal plane design, with each raft containing a 3x3 array of 4k by 4k pixel CCDs. This yields 14 gaps in both the horizontal and vertical direction, for a total loss due to gaps that is close to $5\%$ of the focal plane. Increasing the gaps between chips increases the probability of a detection being lost due to falling in a gap, particularly in the case of the 2-detection cadence, where loss of a single detection results in loss of the entire tracklet.

\section{Results}

We first examine the population of NEOs with $H < 22$ mag, which has been the traditional way of reporting completeness in pursuit of the George E. Brown, Jr. goal. To do this, it was necessary to simulate a population of objects as small as diameters of $70$m, since high albedo objects with that size would have $H\sim22$ mag. 

\begin{figure}[h]
\begin{center}
\includegraphics[width=8cm]{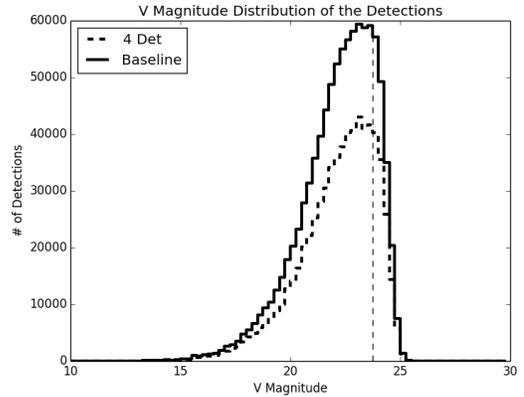}
\caption{The V magnitude distribution of the individual detections in tracks for objects that were successfully found by the baseline and the 4-detection cadence surveys as seen in one of the 25 populations used in this paper. The vertical dashed line indicates a V magnitude of $23.75$ mag. }
\label{fig:vmag_dist}
\end{center}
\end{figure}

The results given in \citet{Ivezic.2007a} and \citet{Ivezic.2008a} are stated in terms of the completeness for the fraction of the population with absolute magnitudes brighter than $H=22$ mag. If we consider the population with $H<22$ mag, we achieve comparable numbers for this population in our simulations. In the 4-detection cadence, our simulations yield a survey completeness of $63\%$ for the PHA population with $H<22$ mag, increasing to $67\%$ completeness for the same population in the 2-detection case. This is $8\%$ lower than the $75\%$ completeness reported in \citet{Ivezic.2007a} and $15\%$ less than the $82\%$ reported in \citet{Ivezic.2008a}. The lower completeness in our simulations can be attributed to the higher fidelity simulations (field-by-field simulation, inclusion of gaps in the focal plane, actual accounting of individual detections needed to form tracklets and tracks), improved knowledge of the population model over the last decade, and better understanding of the performance of the telescope itself. An example of this is the visible magnitude distribution of the detections for the discovered objects in on our simulations as seen in Figure \ref{fig:vmag_dist}. The distribution of $V$ magnitudes turns over sharply at $V \sim 23.5 - 23.75$.  This is inconsistent with Figure 7 in \citet{Myhrvold.2016a}, who assumes $50\%$ limiting magnitudes $\sim0.8$ mag fainter than our computations indicate.

\begin{figure}[h]
\begin{center}
\includegraphics[width=8cm]{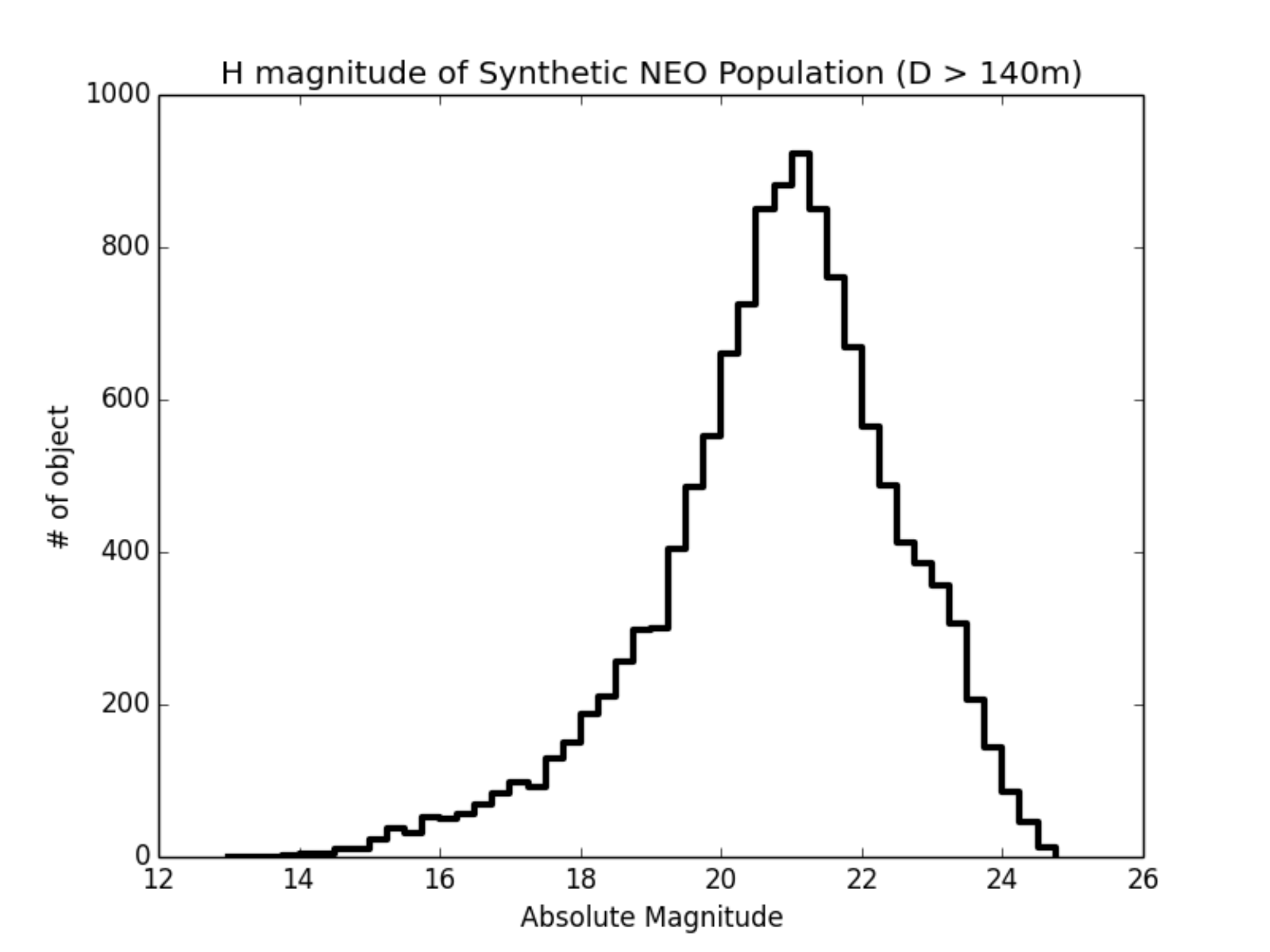}
\caption{The absolute magnitude, $H$, of one of the synthetic NEO populations used in this paper. All the objects have diameters larger than $140$m. About 23\% of the objects have absolute magnitudes fainter than the $H=22$ mag assumed as the default absolute magnitude of an $140$m NEO.}
\label{fig:Hmag_dist}
\end{center}
\end{figure}

However, examination of our improved population models, which now use the size and albedo distributions directly based on the results from the NEOWISE survey \citep{Mainzer.2011e,Mainzer.2012b}, shows that $H < 22$ mag is a poor proxy for the NEO population with diameters larger than $140$m. Approximately one quarter, 23\%, of the NEOs with diameters larger than $140$m have absolute magnitudes fainter than this due to the observed spread in NEO albedos (see Figure \ref{fig:Hmag_dist}). Integrating over the albedo distribution of the NEO population the $90\%$ integral completeness yields $H \sim 22.8$ for the $140$m population.  (E. Wright, personal communication). Setting the target as the population with $H < 22$ mag allows the surveys to discover a large number of high albedo objects with sizes as small as $70$m, while a dark, large object with diameter of $370$m and $2\%$ albedo would be outside the population with $H=22.02$ mag. This effect would be to increase the apparent effectiveness of visible-band surveys, as many objects with diameters less than $140$m would would erroneously be counted in the $H < 22$ mag tally.  Thus, to correctly predict survey completeness, it is necessary to use synthetic populations that properly account for the actual size and albedo distributions of the NEOs, rather than simply assuming $H=22$ mag corresponds to $D=140$m. 

\subsection{Result for the Population of Objects Larger than 140m}

The simulations show that with the 2-detection cadence, LSST reaches a completeness of $63\%$ for the total NEO population larger than $140$m after 10 years, and a $62\%$ completeness for the PHA population larger than $140$m (see Table \ref{tab:lsst_completeness_baseline}).  As expected, LSST is most efficient in finding objects among the Amor subpopulation, due to the fact that much of its time is spent observing the opposition regions of the sky where Amors are preferentially located.  Conversely, LSST is less efficient at finding objects in the Aten and IEO subpopulations, since these tend to be distributed at lower solar elongations, and the two LSST cadences studied here spend relatively little time at solar elongations less than 90$^{\circ}$ (Fig. 1).  High airmass image degradation and phase effects also play a significant role in the reduced detection of IEOs and Atens from ground based observatories.  The completeness falls to $59\%$ for the NEO population with diameters larger than $140$m for the 4-detection cadence tested (see Table \ref{tab:lsst_completeness_4det}). The reduction in completeness is due to the fact that spending more exposures on each patch of sky necessitates a reduction in the amount of fresh sky covered each night. The effect of the 4-detection cadence on the PHA completeness fraction is similar, dropping from $62\%$ to $58\%$.

We have thus computed a more realistic value for the expected completeness of the LSST project in its attempt to satisfy the $140$m goal. We have used the project's most recent simulated survey cadences, which contain the expected limiting magnitude for each field as affected by airmass, sky brightness, and weather. We point to several factors that resulted in a different estimate of the performance of LSST with regards to the NEO and PHA populations. First, we use an improved synthetic model for the NEO and PHA populations, where each object has size, albedo, and orbital elements based on the measured properties of the population from the NEOWISE survey \citep{Mainzer.2011e}. This is a significant improvement over the synthetic population of 800 known PHAs used by \citet{Ivezic.2008a}. Further, our analysis uses diameter to determine the completeness fraction, rather than the absolute magnitudes used in previous work. As pointed out above, a significant fraction of the PHAs with diameters larger than $140$m have absolute magnitudes fainter than the $22$ magnitude limit used in \citet{Ivezic.2008a}. This is due to the fact that as determined by \citet{Mainzer.2011b}, $\sim35\%$ of NEOs have low albedos.  Also, Figure \ref{fig:vmag_dist} shows the V magnitude of the detections of the objects found in the survey simulation for one of the populations used in combination with the baseline {\it enigma\_1189} cadence. It shows that the peak of the distribution is in the $V \sim 23 - 23.75$ mag range, with the distribution turning sharply downwards at $V \sim 23.75$ mag. In both the baseline and the 4-detection cadence, only $18\%$ of the observations of the detected synthetic objects have V magnitudes fainter than $V = 23.75$ mag. This limiting
magnitude is shallower than the $V \sim 24-25$ assumed in \citet{Ivezic.2007a} and \citet{Ivezic.2008a}.

These estimates of completeness assume that LSST would operate without any other surveys, past or future. However, the current NEOs surveys (CSS, Pan-STARRS, NEOWISE, etc.) have already discovered more than 13,000 NEOs, with an estimated current completeness for NEOs larger than $140$m of $\sim25\%$ \citep{Mainzer.2011e}. Simulations of the current surveys estimate that the completeness will rise to $\sim43\%$ at the start of the LSST survey, if the current surveys continue to operate unchanged. Our simulations show that the overlap of objects seen by the current surveys and LSST is significant, with the combination of the LSST and the current surveys at the end of the LSST operations reaching a completeness of $67\%$ for NEOs larger than $140$m (up from the $63\%$ for LSST alone) in the 4-detection cadence case. For the PHAs, the completeness of LSST and the current surveys reaches $71\%$ after the 10 year LSST survey is completed (up from the $58\%$ for LSST alone). This significant overlap is due to the fact that both the current surveys and LSST operate by observing mainly at opposition. For the baseline 2-detection cadence, the combination of LSST and the current surveys reaches $67\%$ and $73\%$ for the NEOs and PHAs larger than $140$m, respectively. 

\begin{figure}[h]
\begin{center}
\includegraphics[width=8cm]{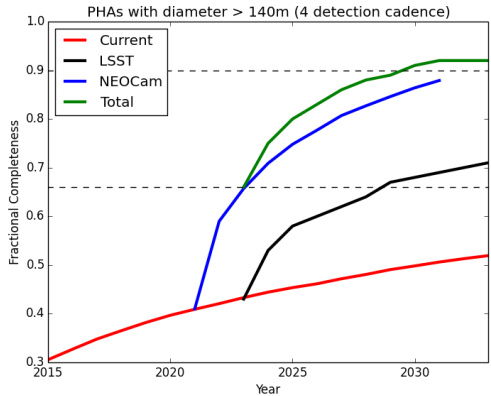}
\caption{The completeness fraction as a function of time based on our simulations of the current surveys (red line), LSST plus the current surveys (black line), and the NEOCam survey plus the current surveys \citep[blue line;][]{Mainzer.2015a}, along with the combined result from these three simulations (green line).  Note that the NEOCam baseline mission is five years; the results of a ten year survey are also shown.}
\label{fig:completeness}
\end{center}
\end{figure}

Figure \ref{fig:completeness} shows the completeness of the PHA population larger than $140$m for the 4-detection cadence. The 25 populations were run through all survey simulations and combined completeness values were derived by counting discovered objects as a function of time as the different surveys become available. Note that we choose to use the 4-detection cadence here, as this is what is used by current surveys and what is most often presented by other studies of survey performance, but using the LSST baseline 2-detection cadence only improves the completeness by a few percent. The performance of the current surveys is shown, along with the combination of the current surveys with the expected LSST. These are compared to the combined completeness of NEOCam and the current surveys, which reach $78\%$ PHA completeness after five years (the NEOCam baseline mission) and $88\%$ by 2031 \citep{Mainzer.2015a}. Operating both LSST and NEOCam offers the fastest means of reaching the $90\%$ goal set by the George E. Brown, Jr. Near-Earth Object section of the NASA Authorization Act of 2005 (Public Law 109-155). 

\begin{deluxetable}{lcccccc}
\tablecolumns{8}
\tablewidth{0pc}
\tablecaption{LSST Yearly Completeness of NEOs $>140$m for 2-detection LSST Cadence}
\tablehead{
\colhead{Year} & \colhead{Total} & \colhead{Interior} & \colhead{Aten} & \colhead{Apollo} & \colhead{Amor} & \colhead{PHAs}  \\
\colhead{} & \colhead{[\%]} & \colhead{[\%]} & \colhead{[\%]} & \colhead{[\%]} & \colhead{[\%]} & \colhead{[\%]}
}
\startdata
1   & $12\pm1$ & $1\pm1$ & $ 9\pm1$ & $12\pm1$ & $14\pm1$ & $13\pm1$ \\
2   & $28\pm1$ & $2\pm1$ & $18\pm1$ & $27\pm1$ & $32\pm1$ & $29\pm1$ \\
3   & $38\pm1$ & $2\pm1$ & $24\pm1$ & $36\pm1$ & $44\pm1$ & $38\pm1$ \\
4   & $45\pm1$ & $2\pm1$ & $30\pm1$ & $43\pm1$ & $52\pm1$ & $44\pm1$ \\
5   & $49\pm1$ & $2\pm1$ & $35\pm2$ & $48\pm1$ & $56\pm1$ & $48\pm1$ \\
6   & $53\pm1$ & $2\pm1$ & $38\pm2$ & $52\pm1$ & $61\pm1$ & $52\pm1$ \\
7   & $56\pm1$ & $2\pm1$ & $40\pm2$ & $55\pm1$ & $64\pm1$ & $55\pm1$ \\
8   & $59\pm1$ & $2\pm1$ & $43\pm2$ & $58\pm1$ & $67\pm1$ & $58\pm1$ \\
9   & $61\pm1$ & $2\pm1$ & $45\pm2$ & $59\pm1$ & $69\pm1$ & $60\pm1$ \\
10 & $63\pm1$ & $2\pm1$ & $47\pm2$ & $61\pm1$ & $70\pm1$ & $62\pm1$ \\
\enddata
\label{tab:lsst_completeness_baseline}
\end{deluxetable}

\begin{deluxetable}{lcccccc}
\tablecolumns{8}
\tablewidth{0pc}
\tablecaption{LSST Yearly Completeness of NEOs $>140$m for 4-detection LSST Cadence}
\tablehead{
\colhead{Year} & \colhead{Total} & \colhead{Interior} & \colhead{Aten} & \colhead{Apollo} & \colhead{Amor} & \colhead{PHAs}  \\
\colhead{} & \colhead{[\%]} & \colhead{[\%]} & \colhead{[\%]} & \colhead{[\%]} & \colhead{[\%]} & \colhead{[\%]}
}
\startdata
1   & $11\pm1$ & $1\pm1$ & $10\pm1$ & $11\pm1$ & $12\pm1$ & $12\pm1$ \\
2   & $25\pm1$ & $1\pm1$ & $15\pm1$ & $24\pm1$ & $30\pm1$ & $26\pm1$ \\
3   & $34\pm1$ & $1\pm1$ & $18\pm1$ & $33\pm1$ & $40\pm1$ & $34\pm1$ \\
4   & $41\pm1$ & $1\pm1$ & $22\pm1$ & $40\pm1$ & $48\pm1$ & $41\pm1$ \\
5   & $46\pm1$ & $1\pm1$ & $25\pm2$ & $45\pm1$ & $54\pm1$ & $46\pm1$ \\
6   & $50\pm1$ & $1\pm1$ & $28\pm2$ & $49\pm1$ & $58\pm1$ & $49\pm1$ \\
7   & $53\pm1$ & $1\pm1$ & $31\pm2$ & $52\pm1$ & $61\pm1$ & $52\pm1$ \\
8   & $55\pm1$ & $1\pm1$ & $33\pm2$ & $55\pm1$ & $64\pm1$ & $55\pm1$ \\
9   & $58\pm1$ & $1\pm1$ & $35\pm2$ & $57\pm1$ & $66\pm1$ & $57\pm1$ \\
10 & $59\pm1$ & $1\pm1$ & $36\pm2$ & $58\pm1$ & $68\pm1$ & $58\pm1$ \\
\enddata
\label{tab:lsst_completeness_4det}
\end{deluxetable}

\section{Discussion and Conclusion}

We have performed a detailed survey simulation of the LSST performance using the current LSST baseline cadence. The simulation shows that if the project is able to reliably generate tracklets using two detections per night and can link these tracklets into a track with a minimum of 3 tracklets covering more than a $\sim6$ day length-of-arc, the survey would discover $62\%$ of the PHAs larger than $140$m in its projected 10 year survey lifetime. This completeness would be reduced to $58\%$ if the 2-detection cadence cannot be implemented, and the more traditional 4-detection cadence is instead adopted. When including the estimated performance from the current operating surveys, assuming these would continue running until the start of LSST and perhaps beyond, the completeness fraction for PHAs larger than $140$m would be $73\%$ for the baseline cadence and $71\%$ for the 4-detection cadence.

Our results differ from the estimates of \citet{Ivezic.2007a} and \citet{Ivezic.2008a}, but are quite comparable to \citet{Jones.2016a} and \citet[][personal communication]{Chesley.2015a}.  Some reasons for the discrepancy include our choice of modeling the survey based on diameter, rather than the proxy population of objects with absolute magnitude $H < 22$ mag; choice of model input populations; and the difference among cadence choices.  Our simulation accounts for the fact that a sizable fraction of NEOs larger than $140$m are dark, with $H>22$ mag \citep{Mainzer.2011b, Mainzer.2012a}. We have shown that using this proxy population is a less than optimal approach for estimating the ability of a survey to make progress towards the George E. Brown, Jr. goal of detecting and tracking $90\%$ of the NEOs larger than $140$m in diameter. 

The advantages of operating both NEOCam and LSST are many.  The combination of LSST and NEOCam creates observational redundancy (which improves reliability of individual tracklets) and the ability to extend orbital arcs, allowing potential impacts to be reliably predicted much farther into the future.  The surveys observe complementary regions of orbital element phase space, with NEOCam observing more interior NEOs and Atens, and LSST preferentially detecting Amors near opposition.  The combination of visible and IR fluxes will produce sizes, albedos, and color information, which gives insight into objects' probable composition \citep[e.g.][]{Mainzer.2012a}.  Composition is a key uncertainty in understanding the potential damage that a given impactor could produce.  

Even if LSST choses to operate in the 2-detection cadence, which may lower its ability to link tracklets, linking its observations to objects found by NEOCam or other sources will provide an immensely powerful capability.  Combining LSST observations with others will help to recover objects, secure orbits, and extend observational arcs.

\section{Acknowledgements}
This publication makes use of data products from NEOWISE, which is a project of the Jet Propulsion Laboratory/California Institute of Technology, funded by the National Aeronautics and Space Administration.  This publication makes use of data products from the \emph{Wide-field Infrared Survey Explorer}, which is a joint project of the University of California, Los Angeles, and the Jet Propulsion Laboratory/California Institute of Technology, funded by the National Aeronautics and Space Administration.  We gratefully acknowledge the support of the JPL High-Performance Computing Facility, which is supported by the JPL Office of the Chief Information Officer.


\end{document}